# Dating of a Latin astrolabe

*Emmanuel Davoust*
*IRAP and Heritage Commission, Observatoire Midi-Pyrénées*

**Abstract**. We have determined the most probable date for the catalog of 34 stars that was used in the construction of a Latin astrolabe originally owned by the Dominican preacher friars and presently at Musée des Arts précieux Paul-Dupuy (Toulouse, France). To this end we digitized a photograph of the rete and the rule of the astrolabe, computed the equatorial coordinates of the ends of the 34 star pointers of the rete, and produced a list of 113 reference stars taken from several lists of stars on astrolabes. We then compared the coordinates of the ends of the pointers and those of the reference stars for dates between 1400 and 1700. The most probable date for this astrolabe is 1550.

## Introduction

The astrolabe is an instrument for making astronomical observations and solving many astronomy problems.

It is made up of a disc (the mater), on the back of which graduations allow you to determine the height of stars over the horizon and the position of the sun in the sky for each day of the year using the alidade. On the front of the mater are successively mounted the plate, the rete and the rule.

The plate is a disk on which curved lines are engraved allowing the altitudes of stars to be converted into hours. As the layout of these lines depends on the latitude of the location of the observer, it is necessary to have a plate for each observation latitude. The present Latin astrolabe has 14 plates.

The rete represents a planar projection of the celestial sphere, with the polar star at its center. It is a cut-out disk comprising:
– a graduated ring (representing the ecliptic circle) on which the names of the twelve constellations of the zodiac are generally engraved,
– a segment of a ring representing the Tropic of Capricorn,
– a horizontal bar showing the intersection of the plane of the celestial equator with that of the ecliptic. This bar cuts the rete into two equal parts, the right ascensions from 0h to 12h below, and those from 12h to 24h above.
– a certain number of pointers (34 for the present astrolabe) whose ends represent the locations of stars.

The rule is a bar for converting lengths into angular distances. It is generally graduated from +50° to -20°. The axis of rotation of the rule corresponds to the declination $\delta = 90°$. Two inscriptions are engraved on the rule of the present astrolabe: "LATITUDO SEPTENTR" and "LAT MERIDIO", and the declination 0° falls between these two inscriptions.

Figure 1 illustrates the values of the right ascension and declination in the different sectors of the rete. The right ascension gradually increases as the rule rotates counterclockwise. The declination, maximum (90°) at the center of the rete, decreases as we move away from the center. It becomes zero on the circle of the celestial equator (not materialized on the rete of the present astrolabe), and negative outside this circle.

The left intersection between the circle of the celestial equator and that of the ecliptic is the vernal point (red point on Figure 1), which is the origin of the equatorial coordinates ($\alpha$ = 0h, $\delta$ = 0° ). The other intersection corresponds to $\alpha$ = 12h, $\delta$ = 0°.

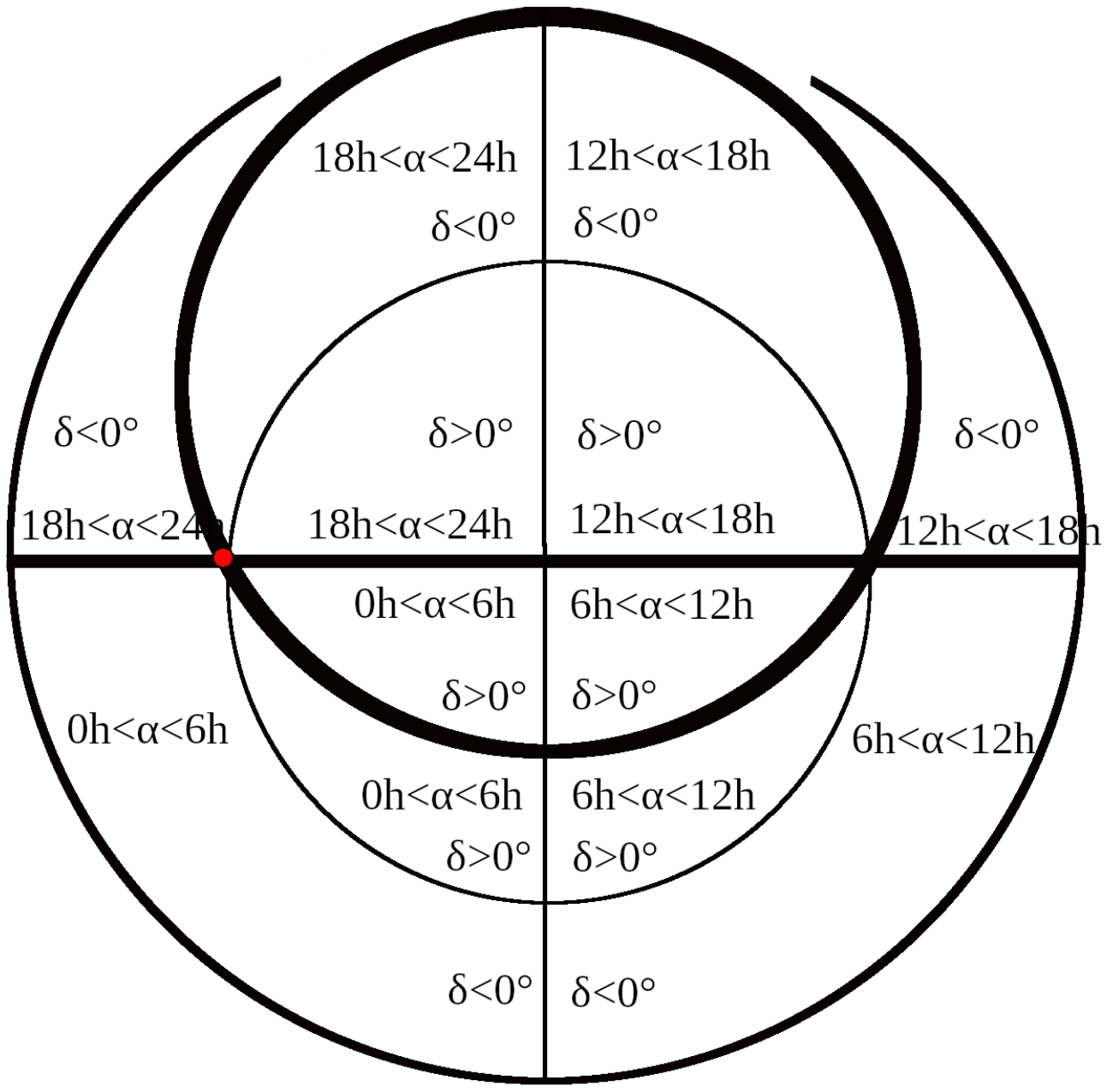

*Figure 1: the different parts of the rete delimit the values of the right ascension α and declination δ*

## Origin of the Latin astrolabe

There are presently two astrolabes at Musée des Arts précieux Paul-Dupuy in Toulouse (France). The Arab astrolabe (inventory number 18088), built at the beginning of the 13th century by Abu Bekr ibn Yusuf, would have belonged to the Ariège astronomer Jacques Vidal (1747-1819), then to Abbot Vidalot-Tornier. The Latin astrolabe studied here comes from a convent of the Dominican preacher friars (probably the convent of Toulouse), as indicated by the inscription *Ord. F.F.P.P*. on the instrument (see Figure 2). It had become the possession of the Toulouse Observatory in Jolimont (inventory number 178) when its director, Benjamin Baillaud, bought it “from Mr. Argance, gilder rue des Chapeliers” on April 19, 1899[1]. It was deposited at the Museum by Observatoire Midi-Pyrénées in July 2005.

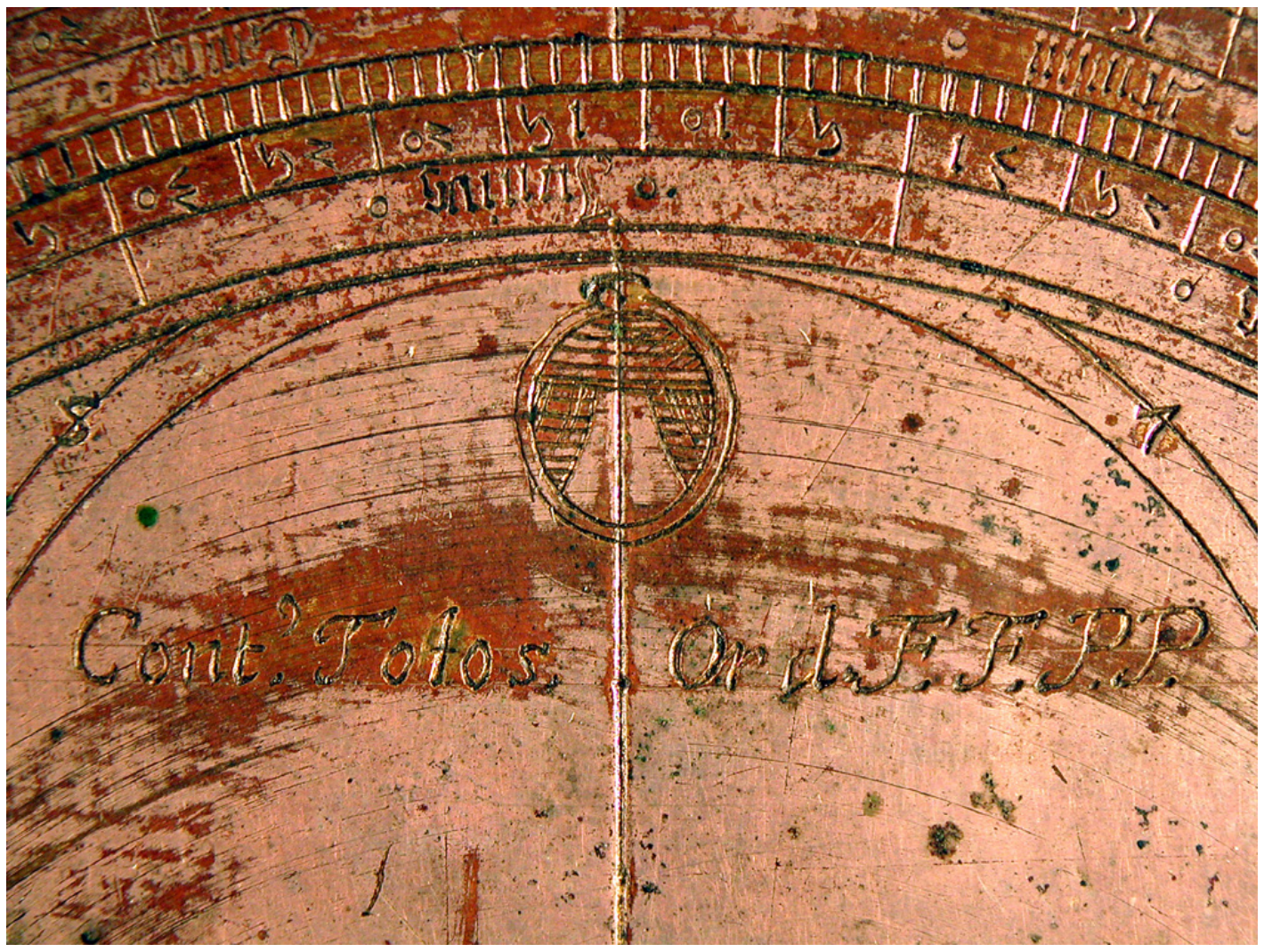

*Figure 2: Inscription on the back of the astrolabe’s mater*

## Dating the astrolabe

There is no date of manufacture on the astrolabe of the Dominican preacher friars, nor are there star names on the rete's 34 pointers. To date the astrolabe, we determined the equatorial coordinates of the

1 Invoice of Mr. Argance to Faculté des Sciences de Toulouse, April 19, 1899. Municipal Archives of Toulouse, inventory number 2R 48.

end of the 34 pointers and produced a list of reference stars, then searched for the reference stars closest to each pointer for different dates.

To determine the equatorial coordinates of the tips of the pointers on the rete, we scanned a photographic print of the rete and the rule. We then measured on the resulting image the positions of the ends of the different pointers and those of the center of the rete in a rectangular coordinate system, as well as the length of the rule in this same system of units (the pixel).

The right ascension $\alpha$ is given by the angle between the rule and the horizontal bar. This angle was determined in two ways, directly with a protractor on the paper print and by calculation from the Cartesian coordinates on the image. The two methods give results in good agreement (at most 0.2 degrees difference) and the second measurement, considered the most precise, was adopted as right ascension.

The rule was used to calculate the declinations of the stars from their geometric coordinates. The declination $\delta$ is obtained by the equation:

$$\delta = \pi/4 - 2*atan(r/l)$$

where $r$ is the distance (in pixels) from the star to the center of the rete and $l$ is the distance (in pixels) between the center of the rule and the graduation 0 on the rule, knowing that $l$ corresponds to a angular distance of 90 degrees in declination. The correspondence between $\delta$ and $l$ is not linear, as shown by the above equation and by the tick marks on the rule which are increasingly tight closer to the axis of rotation.

As the rete does not include star names, we produced a list of bright stars inventoried on astrolabes from the lists given by D'Hollander[2] and Torode[3], from the list of star names on eastern astrolabes from the Adler Planetarium[4], and two lists of astrolabe star names from the National Museum of American History[5]. Unfortunately, we were unable to access the work of Webster et al. (1999) on the Latin astrolabes of the Adler Planetarium. The list of 113 reference stars is given in Table 1 (in the appendix), with the Bayer designation, the equatorial coordinates (sexagesimal format) for the equinox 2000, the proper motions (in milliarcseconds per year), the B magnitude, the frequency of occurrence of the star name in the different lists. Note that *alpha Andromedae* is the same star as *delta Pegasi* and *beta Tauri* is the same as *gamma Aurigae*.

Although they constitute the sphere of “fixed stars”, stars do not have time-independent coordinates. These coordinates change regularly because of the precession of the equinoxes, a consequence of the slow change in direction of the earth's axis of rotation. Additionally, stars have small relative displacements with respect to each other, called proper motions. It is precisely these changes in coordinates over time that make it possible to date a catalog of stars.

We extracted the equatorial coordinates for the equinox 2000 and the proper motions of the stars of the reference list from the SIMBAD[6] online database. Then we calculated the coordinates of these stars for all equinoxes between 1400 and 1700 with a step of 50, using the precession equations given in the Astronomical Almanac for 2005, p.B18. We took into account the proper motions of the stars in these calculations, although the effect of these displacements is small compared to the uncertainties of the coordinates on the astrolabe. Finally we compared these coordinates with those of the ends of the 34 point-

2 Tables 3.16, 3.17 and 12.4 of D'Hollander 1999
3 Table 4 of Torode 1992
4 P. 244-247 of Pingree et al. 2009
5 Tables IIA and IID of Gibbs & Saliba 1984
6 https://simbad.u-strasbg.fr/simbad/

ers for each equinox and found the most probable equinox, 1550, by minimizing the difference between the coordinates of the pointer ends and those of the associated stars. We tried to improve the agreement between the position of the reference stars and that of the pointer ends by slightly modifying the coordinates of the center of the rete, or the length of the rule, or by using the 1570 and 1530 equinoxes. The 1550 equinox remains the one that gives the best agreement.

As two pointer ends (11 and 16) have no matches in our reference list, we searched for other possible candidates in the SIMBAD database. To do this, we limited ourselves to stars brighter than magnitude 5 and with proper motions. We calculated the coordinates of these stars for the 1550 equinox. We were thus able to find two stars not used by astrolabists, *16 Lyncis* which is in poor agreement with pointer 11 and *phi Leonis*, which is in poor agreement with pointer 16.

Table 2 (in the appendix) gives the number of the pointer, the equatorial coordinates (decimal format) of the pointer, as well as the coordinates for the 1550 equinox, separation in degrees, name and magnitude of the star associated with this pointer. A v in front of the magnitude indicates a variable star. The last columns give the number of astrolabes on which the star was identified, successively on the eastern astrolabes in the work of Pingree et al. (2009), the Arabic and Latin astrolabes in the work of Smith and Saliba (1984) and those in the article by Torode (1992), as well as the mention straight or broken pointer, where applicable. In a few cases we have indicated several stars as possible.

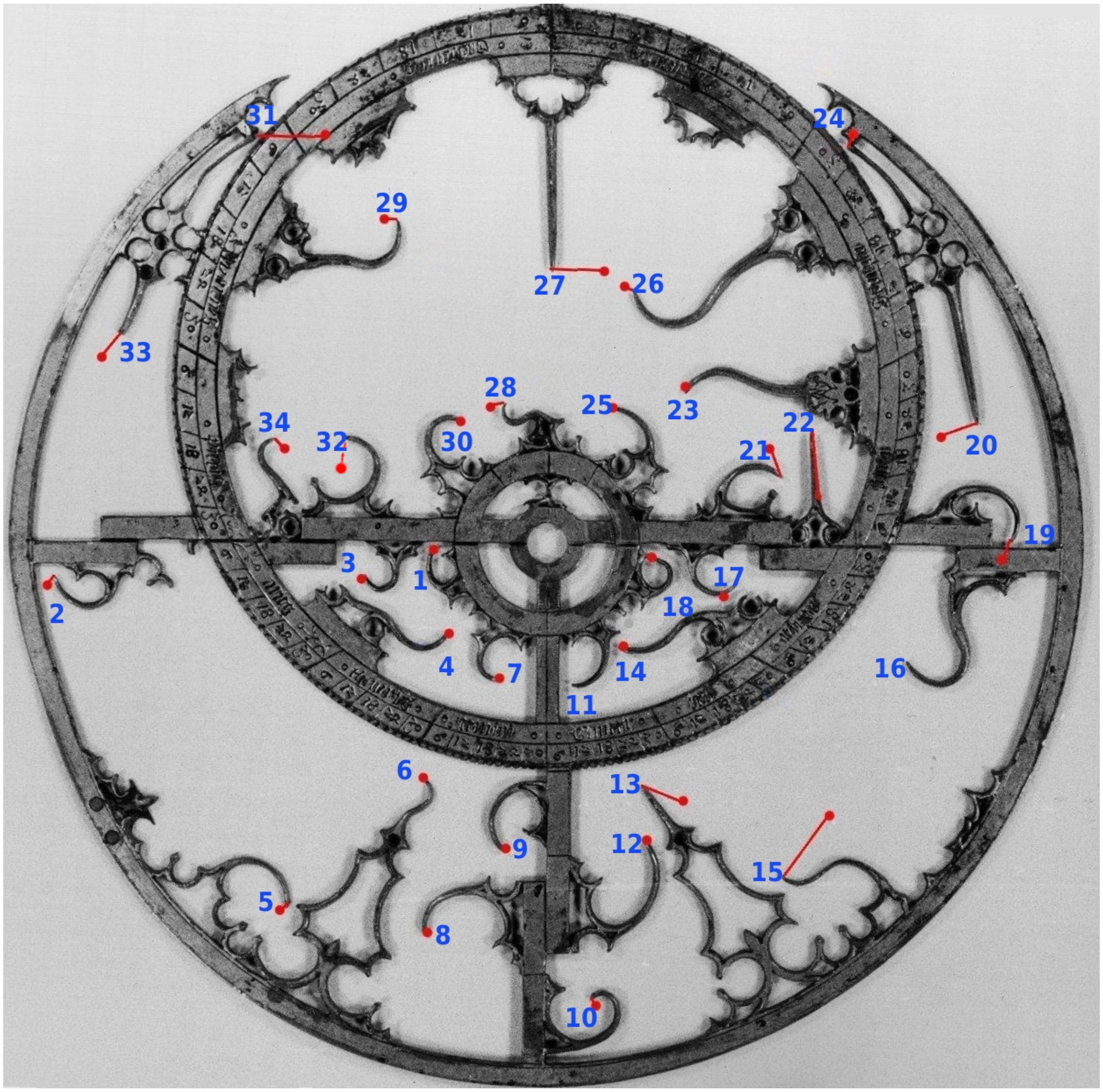

*Figure 3: the rete, pointers (blue numbers), reference stars (red dots)*

Figure 3 shows the qualitative agreement between the ends of the pointers (numbered in blue) and the associated stars (red dot), which in 23 of 34 cases is very satisfactory, the angular separation being less than three degrees. In four cases (13, 20, 22, 27), the pointer is straight and the agreement poor, suggesting that the astrolabe remained unfinished. In one case (33), the pointer is broken, and a bright star is in its extension. In summary, there are only two pointers (11, 16) without a star used by other astrolabes, and to which we could not associate a star from the reference list.

We assume that errors or inaccuracies in the catalog of stars used in the construction of this astrolabe are at the origin of the deviations observed and the failure for pointers 11 and 16.

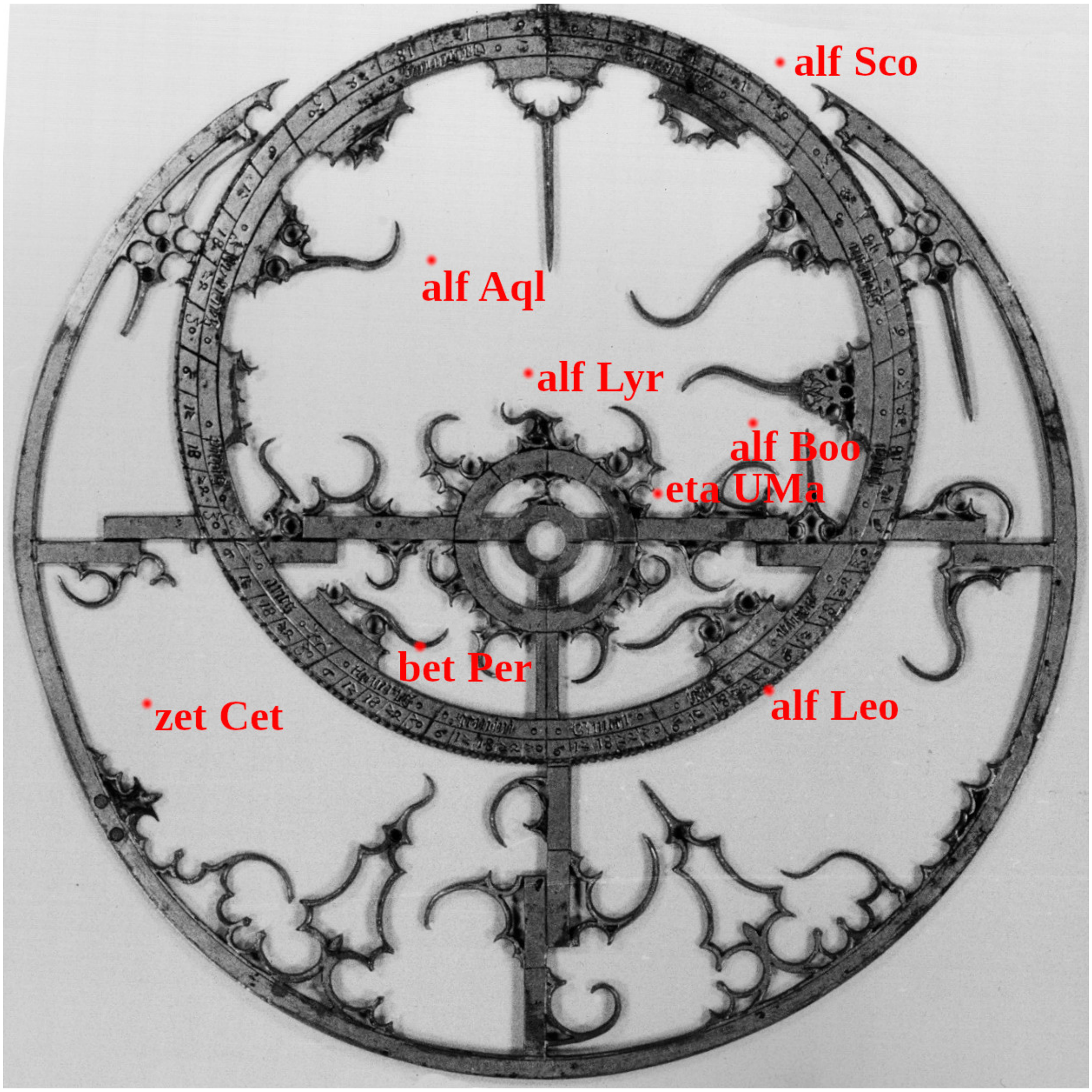

*Figure 4: Positions of the 8 stars of Torode's (1992) list absent from the astrolabe's star catalog*

Figure 4 shows the location of the stars from Torode's (1992) list that were not used for this astrolabe, i.e. 8 stars out of 20. We recall that this list gives the names of the 20 stars represented on 74 or more astrolabes studied by the author. The star *bet Per* (Algol) would have given a poorer agreement with the pointer 4 than the brighter star *alf Per* (Mirfak). The stars *alf Aql*, *alf Lyr* and *alf Boo*, although very bright, are more than 10 degrees from the nearest pointer. The other four stars could not have been part of the catalog without the addition of pointers to the spider.

## Peculiarities of the astrolabe

This astrolabe is not signed. As we dated to 1550 the star catalog which was used to construct it, we could attribute its construction to a member of the Flemish school of astrolabists of that time, in particular to the itinerant astrolabist Adrien Descrolières (active in the second half of the 16th century). It turns out that an unsigned astronomical quadrant by Descrolières, also from the convent of the Dominican preacher friars of Toulouse, is currently in Musée des Arts précieux Paul-Dupuy. But the rete of that quadrant looks nothing like the one of our astrolabe. Another possibility would be Gautier Arsenius, who was active in Louvain between 1561 and 1575. Note that 19 of the 34 stars of our astrolabe are identified on the Arsenius astrolabe of the Institut Géographique National, which has a total of 41 stars (D' Hollander 1999).

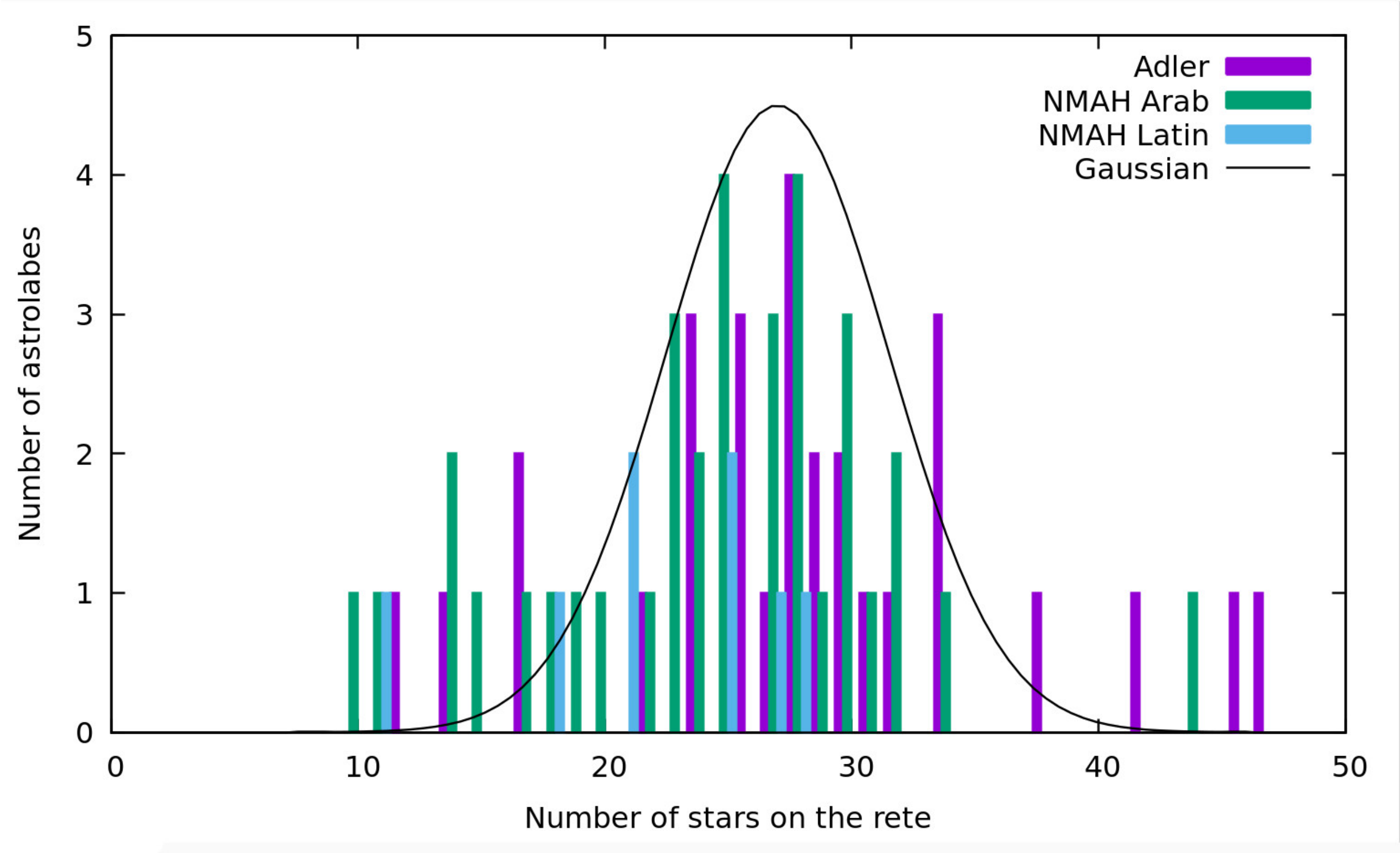

*Figure 5: Histogram of the number of stars on astrolabes in three lists of astrolabe stars*

There are 34 star pointers on the astrolabe of the Dominical preacher friars. This is a relatively large number compared to the number on the various astrolabes described in the two major monographs, those of Pingree et al. (2009) and Smith and Saliba (1984): in the Adler Planetarium collection, 4 out of 29 astrolabes have more than 33 stars. In that of the National Museum of American History, the figure is 1 on 35 Arabic astrolabes and 0 on 8 Latin astrolabes. The Spanish astrolabe studied by King (2003) has 21 stars. The histogram on Figure 5, established from three lists of astrolabe stars, shows that an astrolabe has 27 stars on average. The Gaussian fitted to the histogram has a standard deviation of 4.5. The difference in the number of stars in our astrolabe from the average is therefore significant.

## Acknowledgments

We thank Jacques Cadaugade who photographed the astrolabe and digitized the photographic prints. This research has made use of the SIMBAD database, operated at CDS, Strasbourg, France. This is the translation of a technical report (in French) for the Musée des Arts précieux Paul-Dupuy, which is available at https://hal.science/hal-04269582v2

## Appendix

Table 1 : reference list of stars

| Bayer designation | coordinates J2000 | Proper motions | | mag B | A,Sa,Sl, | To |
|---|---|---|---|---|---|---|
| alpha Andromedae | 00 08 23.260 +29 05 25.55 | 137.46 | -163.44 | 2.012 | 0,3,0 | |
| beta Andromedae | 01 09 43.924 +35 37 14.01 | 175.90 | -112.20 | 3.64 | 6,2,3 | |
| gamma Andromedae | 02 03 53.953 +42 19 47.01 | 43.08 | -50.85 | 3.550 | 3,2,0 | |
| alpha Aquarii | 22 05 47.036 -00 19 11.46 | 18.25 | -9.39 | 3.905 | 0,0,3 | |
| delta Aquarii | 22 54 39.014 -15 49 14.98 | -42.60 | -27.89 | 3.344 | 16,20,2 | |
| alpha Aquilae | 19 50 46.999 +08 52 05.96 | 536.23 | 385.29 | 0.99 | 29,34,5 | 131 |
| zeta Aquilae | 19 05 24.608 +13 51 48.52 | -7.25 | -95.56 | 3.016 | 4,3,0 | |
| theta Aquilae | 20 11 18.266 -00 49 17.31 | 35.26 | 5.71 | 3.197 | 0,0,0 | |
| xi Aquilae | 19 54 14.882 +08 27 41.23 | 101.91 | -81.20 | 5.769 | 0,0,0 | |
| beta Arietis | 01 54 38.411 +20 48 28.91 | 98.74 | -110.41 | 2.77 | 1,0,0 | |
| gamma Arietis | 01 53 31.815 +19 17 37.88 | 79.20 | -97.63 | 3.84 | 1,0,0 | |
| delta Arietis | 03 11 37.765 +19 43 36.04 | 153.33 | -8.28 | 5.376 | 0,1,0 | |
| alpha Aurigae | 05 16 41.359 +45 59 52.77 | 75.25 | -426.89 | 0.88 | 25,21,6 | 114 |
| beta Aurigae | 05 59 31.723 +44 56 50.76 | -56.44 | -0.95 | 1.969 | 0,1,0 | |
| alpha Bootis | 14 15 39.672 +19 10 56.67 | -1093.39 | -2000.06 | 1.19 | 12,34,8 | 132 |
| beta Bootis | 15 01 56.762 +40 23 26.04 | -40.15 | -28.86 | 4.431 | 0,1,0 | |
| gamma Bootis | 14 32 04.672 +38 18 29.70 | -115.71 | 151.16 | 3.23 | 0,0,2 | |
| epsilon Bootis | 14 44 59.217 +27 04 27.21 | -50.95 | 21.07 | 3.36 | 0,0,1 | |
| eta Bootis | 13 54 41.079 +18 23 51.79 | -60.95 | -356.29 | 3.250 | 0,0,0 | |
| theta Bootis | 14 25 11.797 +51 51 02.68 | -235.40 | -399.07 | 4.55 | 17,0,0 | |
| alpha Cancri | 08 58 29.223 +11 51 27.72 | 43.23 | -29.63 | 4.401 | 1,2,0 | |
| beta Cancri | 08 16 30.921 +09 11 07.96 | -46.82 | -49.24 | 5.000 | 0,1,0 | |
| epsilon Cancri | 08 40 27.011 +19 32 41.31 | -35.60 | -12.98 | 6.460 | 0,1,0 | |
| alpha Canis maioris | 06 45 08.917 -16 42 58.02 | -546.01 | -1223.07 | -1.46 | 24,34,9 | 137 |
| zeta Canis maioris | 06 20 18.792 -30 03 48.12 | 7.32 | 4.03 | 2.83 | 1,0,0 | |
| alpha Canis minoris | 07 39 18.119 +05 13 29.96 | -714.59 | -1036.80 | 0.74 | 27,32,7 | 128 |
| beta Canis minoris | 07 27 09.042 +08 17 21.54 | -51.76 | -38.29 | 2.814 | 0,2,0 | |
| beta Capricorni | 20 21 00.673 -14 46 52.98 | 44.92 | 7.38 | 3.87 | 1,0,0 | |
| gamma Capricorni | 21 40 05.456 -16 39 44.31 | 187.56 | -22.45 | 4.00 | 1,0,0 | |
| delta Capricorni | 21 47 02.444 -16 07 38.23 | 261.70 | -296.70 | 3.16 | 15,10,2 | |
| alpha Carinae | 06 23 57.110 -52 41 44.38 | 19.93 | 23.24 | -0.57 | 0,0,0 | |
| alpha Cassiopeiae | 00 40 30.441 +56 32 14.39 | 50.88 | -32.13 | 3.434 | 0,0,1 | |
| beta Cassiopeiae | 00 09 10.685 +59 08 59.21 | 523.50 | -179.77 | 2.61 | 24,21,0 | |
| alpha Ceti | 03 02 16.773 +04 05 23.06 | -10.41 | -76.85 | 4.15 | 2,5,2 | |
| beta Ceti | 00 43 35.371 -17 59 11.78 | 232.55 | 31.99 | 3.06 | 11,21,6 | |
| gamma Ceti | 02 43 18.039 +03 14 08.94 | -146.10 | -146.12 | 3.56 | 13,23,0 | |
| zeta Ceti | 01 51 27.635 -10 20 06.13 | 40.80 | -37.25 | 4.873 | 5,3,8,98 | |
| iota Ceti | 00 19 25.674 -08 49 26.11 | -15.15 | -37.11 | 4.782 | 21,7,0 | |
| pi Ceti | 02 44 07.349 -13 51 31.31 | -8.62 | -9.07 | 4.123 | 15,0,0 | |
| tau Ceti | 01 44 04.083 -15 56 14.93 | -1721.05 | 854.16 | 4.22 | 1,0,0 | |
| alpha Coronae borealis | 15 34 41.268 +26 42 52.89 | 120.27 | -89.58 | 2.244 | 28,33,3 | 107 |
| alpha Corvi | 12 08 24.817 -24 43 43.95 | 99.52 | -39.19 | 4.34 | 0,0,1 | |
| gamma Corvi | 12 15 48.371 -17 32 30.95 | - 158.61 | 21.86 | 2.48 | 23,21,3 | |
| alpha Crateris | 10 59 46.465 -18 17 55.62 | -462.26 | 129.49 | 5.17 | 17,17,2 | |
| alpha Cygni | 20 41 25.915 +45 16 49.22 | 2.01 | 1.85 | 1.34 | 24,28,5 | 96 |
| beta Cygni | 19 30 43.281 +27 57 34.85 | -7.17 | -6.15 | 4.171 | 6,8,0 | |
| delta Cygni | 19 44 58.479 +45 07 50.92 | 44.07 | 48.66 | 2.881 | 0,0,0 | |
| epsilon Cygni | 20 46 12.682 +33 58 12.92 | 355.66 | 330.60 | 3.520 | 0,0,0 | |
| epsilon Delphini | 20 33 12.772 +11 18 11.74 | 11.96 | -28.97 | 3.920 | 13,6,3 | |
| alpha Draconis | 14 04 23.350 +64 22 33.06 | -56.34 | 17.21 | 3.640 | 4,0,0 | |
| gamma Draconis | 17 56 36.370 +51 29 20.02 | -8.48 | -22.79 | 3.760 | 0,1,2 | |
| gamma Eridani | 03 58 01.767 -13 30 30.67 | 61.57 | -113.11 | 4.595 | 0,11,0 | |
| theta Eridani | 02 58 15.675 -40 18 16.85 | -52.89 | 21.98 | 3.03 | 17,,0 | |
| tau Eridani | 02 45 06.187 -18 34 21.21 | 334.20 | 37.19 | 4.95 | 0,1,0 | |
| alpha Geminorum | 07 34 35.873 +31 53 17.82 | -191.45 | -145.19 | 1.63 | 0,2,1 | |
| alpha Herculis | 17 14 38.858 +14 23 25.23 | -7.32 | 36.07 | 4.51 | 6,2,2 | |
| sigma Herculis | 16 34 06.183 +42 26 13.35 | -7.54 | 59.42 | 4.187 | 0,0,0 | |
| tau Herculis | 16 19 44.437 +46 18 48.11 | -13.33 | 38.48 | 3.750 | 0,0,0 | |
| alpha Hydrae | 09 27 35.243 -08 39 30.96 | -15.23 | 34.37 | 3.486 | 27,29,6 | 107 |
| sigma Hydrae | 08 38 45.437 +03 20 29.17 | -19.48 | -15.92 | 5.678 | 3,0,0 | |
| alpha Leonis | 10 08 22.311 +11 58 01.95 | -248.73 | 5.59 | 1.24 | 14,19,7 | 122 |
| beta Leonis | 11 49 03.578 +14 34 19.41 | -497.68 | -114.67 | 2.23 | 1,6,2 | |
| delta Leonis | 11 14 06.501 +20 31 25.39 | 143.42 | -129.88 | 2.68 | 2,0,0 | |
| epsilon Leonis | 09 45 51.073 +23 46 27.32 | -45.61 | -9.21 | 3.762 | 2,0,0 | |
| alpha Librae | 14 50 52.713 -16 02 30.40 | -105.68 | -68.40 | 2.912 | 1,0,1 | |
| beta Librae | 15 17 00.414 -09 22 58.49 | -98.10 | -19.65 | 2.535 | 2,0,0 | |
| alpha Lyrae | 18 36 56.336 +38 47 01.28 | 200.94 | 286.23 | 0.03 | 29,33,8 | 132 |
| alpha Ophiuchi | 17 34 56.069 +12 33 36.13 | 108.07 | -221.57 | 2.23 | 25,31,5 | 95 |
| delta Ophiuchi | 16 14 20.739 -03 41 39.56 | -47.54 | -142.73 | 4.32 | 3,0,2 | |
| zeta Ophiuchi | 16 37 09.539 -10 34 01.53 | 15.26 | 24.79 | 2.595 | 3,2,0 | |
| nu Ophiuchi | 17 59 01.592 -09 46 25.08 | -9.48 | -116.69 | 4.31 | 0,0,0 | |
| alpha Orionis | 05 55 10.305 +07 24 25.43 | 27.54 | 11.30 | 2.27 | 22,27,3 | 95 |
| beta Orionis | 05 14 32.272 -08 12 05.90 | 1.31 | 0.50 | 0.09 | 26,30,6 | 125 |
| gamma Orionis | 05 25 07.863 +06 20 58.93 | -8.11 | -12.88 | 1.42 | 8,4,0 | |
| kappa Orionis | 05 47 45.389 -09 40 10.58 | 1.46 | -1.28 | 1.937 | 9,0,0 | |
| chi Orionis | 05 54 22.983 +20 16 34.22 | -162.54 | -99.51 | 5.00 | 0,4,2 | |
| alpha Pegasi | 23 04 45.653 +15 12 18.96 | 60.40 | -41.30 | 2.45 | 2,0,6 | |

| | | | | | | | |
|---|---|---|---|---|---|---|---|
| beta Pegasi | 23 03 46.457 | +28 04 58.03 | 187.65 | 136.93 | 4.09 | 26,22,3 | 104 |
| gamma Pegasi | 00 13 14.151 | +15 11 00.94 | 1.98 | -9.28 | 2.60 | 1,2,0 | |
| epsilon Pegasi | 21 44 11.156 | +09 52 30.03 | 26.92 | 0.44 | 3.962 | 18,21,1 | |
| kappa Pegasi | 21 44 38.735 | +25 38 42.14 | 48.13 | 14.29 | 4.550 | 0,0,0 | |
| tau Pegasi | 23 20 38.242 | +23 44 25.21 | 29.45 | -9.53 | 4.756 | 0,1,0 | |
| alpha Persei | 03 24 19.370 | +49 51 40.25 | 23.75 | -26.23 | 2.286 | 0,2,3 | |
| beta Persei | 03 08 10.132 | +40 57 20.33 | 2.99 | -1.66 | 2.07 | 15,21,3 | 79 |
| omicron Persei | 03 44 19.132 | +32 17 17.69 | 8.18 | -10.43 | 3.871 | 1,0,0 | |
| alpha Piscis austrini | 22 57 39.046 | -29 37 20.05 | 328.95 | -164.67 | 1.25 | 0,1,0 | |
| zeta Puppis | 08 03 35.048 | -40 00 11.33 | -29.71 | 16.68 | 1.97 | 1,0,0 | |
| rho Puppis | 08 07 32.649 | -24 18 15.57 | -83.35 | 46.23 | 3.24 | 4,2,1 | |
| rho Sagitarii | 19 21 40.359 | -17 50 49.92 | -25.87 | 21.46 | 4.149 | 2,0,0 | |
| alpha Scorpii | 16 29 24.460 | -26 25 55.21 | -12.11 | -23.30 | 2.96 | 18,23,3 | 85 |
| beta Scorpii | 16 05 26.232 | -19 48 19.63 | -5.20 | -24.04 | 2.55 | 0,0,0 | |
| delta Scorpii | 16 00 20.005 | -22 37 18.14 | -10.21 | -35.41 | 2.205 | 0,0,0 | |
| lambda Scorpii | 17 33 36.520 | -37 06 13.76 | -8.53 | -30.80 | 1.48 | 0,0,0 | |
| alpha Serpentis | 15 44 16.074 | +06 25 32.26 | 133.84 | 44.81 | 3.800 | 23,24,0 | |
| rho Serpentis | 15 51 15.910 | +20 58 40.52 | -53.32 | 18.87 | 6.301 | 1,0,0 | |
| alpha Tauri | 04 35 55.239 | +16 30 33.49 | 63.45 | -188.94 | 2.468 | 24,33,8 | 133 |
| beta Tauri | 05 26 17.513 | +28 36 26.83 | 22.76 | -173.58 | 1.62 | 0,2,0 | |
| eta Tauri | 03 47 29.077 | +24 06 18.49 | 19.34 | -43.67 | 2.806 | 0,0,1 | |
| alpha Trianguli | 01 53 04.907 | +29 34 43.78 | 10.82 | -234.24 | 3.90 | 0,3,0 | |
| alpha Ursae maioris | 11 03 43.672 | +61 45 03.72 | -134.11 | -34.70 | 2.86 | 17,12,3 | |
| beta Ursae maioris | 11 01 50.477 | +56 22 56.73 | 81.43 | 33.49 | 2.376 | 0,0,1 | |
| gamma Ursae maioris | 11 53 49.847 | +53 41 41.14 | 107.68 | 11.01 | 2.450 | 0,0,0 | |
| delta Ursae maioris | 12 15 25.561 | +57 01 57.42 | 104.11 | 7.30 | 3.410 | 0,0,0 | |
| epsilon Ursae maioris | 12 54 01.750 | +55 57 35.36 | 111.91 | -8.24 | 1.801 | 1,0,1 | |
| zeta Ursae maioris | 13 23 55.540 | +54 55 31.27 | 119.01 | -25.97 | 2.29 | 6,2,0 | |
| eta Ursae maioris | 13 47 32.438 | +49 18 47.76 | -121.17 | -14.91 | 1.755 | 6,3,3 | 74 |
| theta Ursae maioris | 09 32 51.434 | +51 40 38.28 | -947.46 | -535.60 | 3.63 | 0,0,0 | |
| iota Ursae maioris | 08 59 12.454 | +48 02 30.57 | -441.29 | -215.32 | 3.33 | 0,0,1 | |
| mu Ursae maioris | 10 22 19.740 | +41 29 58.27 | -81.47 | 35.34 | 4.672 | 5,2,0 | |
| xi Ursae maioris | 11 18 10.9 | +31 31 44 | -430.00 | -588.00 | 4.38 | 0,0,0 | |
| alpha Ursae minoris | 02 31 49.095 | +89 15 50.79 | 44.48 | -11.85 | 2.591 | 0,2,0 | |
| alpha Virginis | 13 25 11.579 | -11 09 40.75 | -42.35 | -30.67 | 0.91 | 27,33,7 | 129 |
| epsilon Virginis | 13 02 10.598 | +10 57 32.94 | -273.80 | 19.96 | 3.77 | 1,2,0 | |

Tableau 2 : catalog of the astrolabe's stars

| **Np** | **RAp** | **Decp** | **RA*** | **Dec*** | **Sep** | **Name** | **magB** | **A,Sa,Sl** | **To** | **pointer** |
|---|---|---|---|---|---|---|---|---|---|---|
| 1 | 0h14 | 54d57 | 0h15 | +54°03 | 0.9 | alpha Cassiopeiae | v2.2 | 0, 0, 1 | | |
| 2 | 0h16 | -19°33 | 0h20 | -20°28 | 1.3 | beta Ceti | 2.0 | 11,21, 6 | | |
| 3 | 0h40 | 34°23 | 0h45 | +33°12 | 1.6 | beta Andromedae | 2.0 | 6, 2, 3 | | |
| 4 | 2h47 | 48°32 | 2h53 | +48°09 | 1.1 | alpha Persei | 1.8 | 0, 2, 3 | | |
| 5 | 3h39 | -14°19 | 3h37 | -14°51 | 0.7 | gamma Eridani | 3.0 | 0,11, 0 | | |
| 6 | 4h11 | 15°27 | 4h10 | +15°30 | 0.2 | alpha Tauri | 0.8 | 24,33, 8 | 133 | |
| 7 | 4h44 | 45°19 | 4h43 | +45°24 | 0.2 | alpha Aurigae | 0.1 | 25,21, 6 | 114 | |
| 8 | 4h52 | -09°21 | 4h53 | -08°48 | 0.6 | beta Orionis | 0.2 | 26,30, 6 | 125 | |
| 9 | 5h31 | 6°21 | 5h30 | +07°13 | 0.9 | alpha Orionis | 0.8 | 22,27, 3 | 95 | |
| 10 | 6h23 | -15°30 | 6h25 | -16°10 | 0.8 | alpha Canis maioris | -1.4 | 24,34, 9 | 137 | |
| 11 | 6h40 | 45°16 | 6h24 | +45°32 | 2.8 | 16 Lyncis | 4.9 | | | |
| 12 | 7h13 | 6°35 | 7h16 | +06°17 | 0.8 | alpha Canis minoris | 0.4 | 27,32, 7 | 128 | |
| 13 | 7h26 | 16°35 | 7h52 | +10°28 | 8.8 | beta Cancri | 3.5 | 0, 1, 0 | | straight |
| 14 | 8h18 | 50°19 | 8h28 | +49°42 | 1.7 | iota Ursa maioris | 3.2 | 0, 0, 1 | | |
| 15 | 8h21 | -08°55 | 9h05 | -06°46 | 11.1 | alpha Hydrae | 2.0 | 27,29, 6 | 107 | |
| 16 | 10h48 | -04°31 | 10h54 | -01°12 | 3.6 | phi Leonis | 4.5 | | | |
| 17 | 11h14 | 31°47 | 10h54 | +34°02 | 4.8 | ksi UMa | 2.3 | 0, 0, 0 | | |
| 18 | 11h23 | 56°29 | 11h29 | +56°12 | 0.9 | gamma Ursa Maioris | 2.4 | 0, 0, 0 | | |
| 19 | 12h 1 | -15°57 | 11h52 | -15°02 | 2.4 | gamma Corvi | 2.6 | 23,21, 3 | | |
| 20 | 13h 2 | -13°50 | 13h02 | -08°46 | 5.1 | alpha Virginis | v1.0 | 27,33, 7 | 129 | straight |
| 21 | 13h 2 | 19°57 | 13h33 | +20°41 | 7.3 | eta Bootis | 2.7 | 0, 0, 0 | | |
| 22 | 13h33 | 11° 3 | 12h39 | +13°24 | 13.4 | epsilon Virginis | 2.8 | 1, 2, 0 | | |
| 23 | 15h10 | 29°38 | 15h16 | +28°17 | 1.9 | alpha Coronae borealis | v2.3 | 28,33, 3 | 107 | |
| 24 | 15h33 | -19°56 | 15h34 | -21°14 | 1.3 | delta Scorpii | 2.3 | 0, 0, 0 | | |
| 24 | | | 15h33 | -21°14 | 1.3 | beta Scorpii | 2.6 | 0, 0, 0 | | |
| 25 | 16h18 | 43°40 | 16h19 | +43°25 | 0.3 | sigma Herculis | 4.2 | 0, 0, 0 | | |
| 26 | 16h44 | 15°14 | 16h54 | +14°59 | 2.4 | alpha Herculis | 3.1 | 6, 2, 2 | | |
| 27 | 17h56 | 14°30 | 17h14 | +12°57 | 10.3 | alpha Ophiuchi | 2.1 | 25,31,5 | 95 | straight |
| 28 | 19h 6 | 44°43 | 19h31 | +44°05 | 4.5 | delta Cygni | 2.9 | 0, 0, 0 | | |
| 29 | 19h41 | -01°13 | 19h48 | -02°04 | 1.9 | teta Aquilae | 3.2 | 0, 0, 0 | | |
| 30 | 20h20 | 43d47 | 20h26 | +43d43 | 1.1 | alpha Cygni | 1.2 | 24,28, 5 | 96 | |
| 31 | 20h24 | -20° 6 | 20h03 | -19°12 | 5.0 | rho Capricorni | 4.8 | 0, 0, 0 | | straight |
| 31 | | | 20h19 | -26°49 | 6.8 | psi Capricorni | 4.1 | 0, 0, 0 | | |
| 31 | | | 19h56 | -16°06 | 7.8 | beta Capricorni | 3.1 | 0, 0, 0 | | |
| 32 | 22h13 | 23° 3 | 22h22 | +27°54 | 5.3 | eta Pegasi | 3.8 | 0, 0, 0 | | |
| 33 | | | 22h42 | +25°40 | 7.1 | beta Pegasi | v2.4 | 26,22, 3 | 104 | |
| 33 | 22h17 | -17°20 | 22h31 | -18°11 | 3.4 | delta Aquarii | 3.3 | 16,20, 2 | | broken |
| 34 | 22h37 | 10°10 | 22h42 | +12°48 | 2.9 | alpha Pegasi | 2.5 | 2, 0, 6 | | |

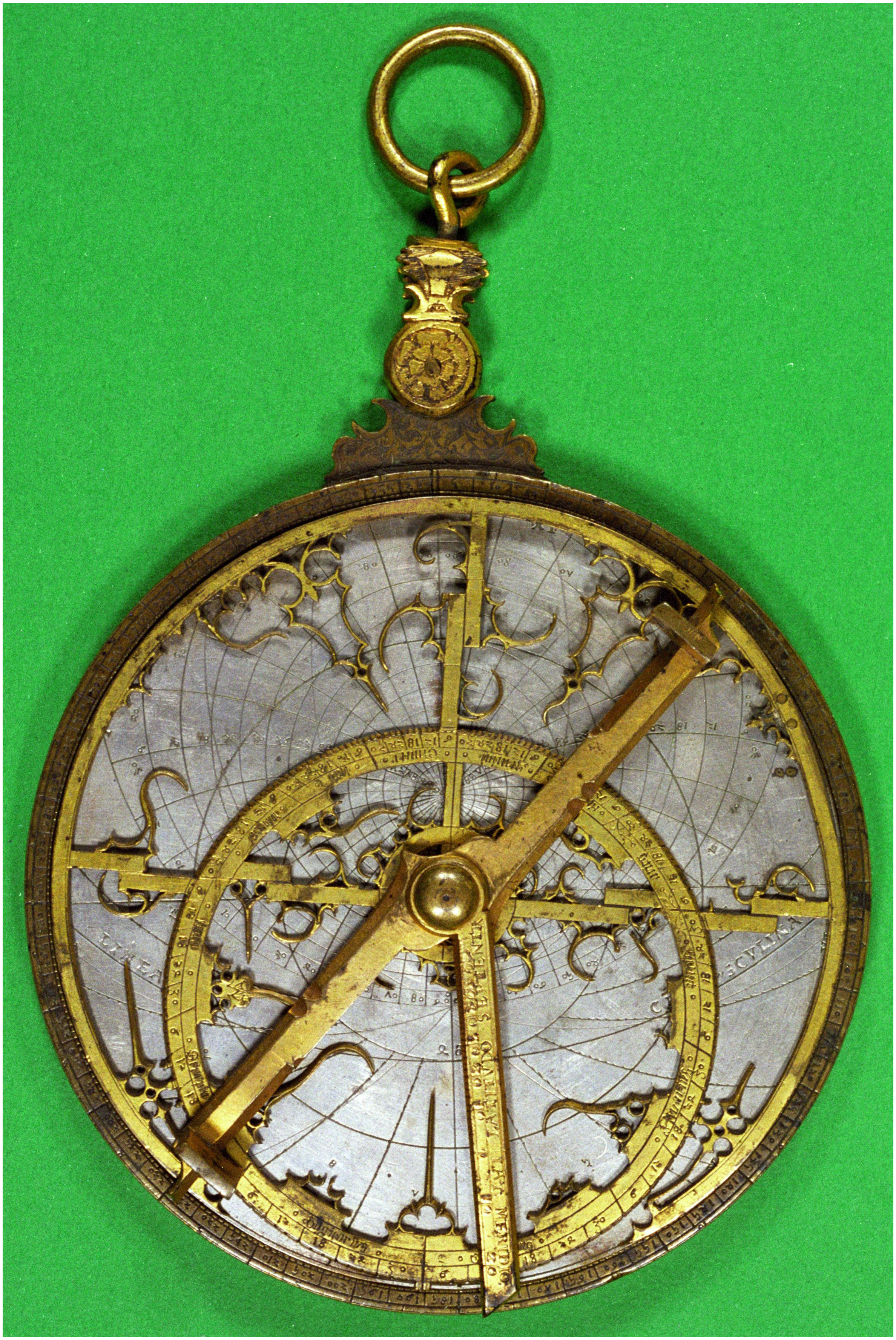

*Figure 6: The front of the astrolabe, with a plate, the rete, the alidade (diagonal bar), the rule (vertical bar). The alidade is in principle mounted on the back of the mater*

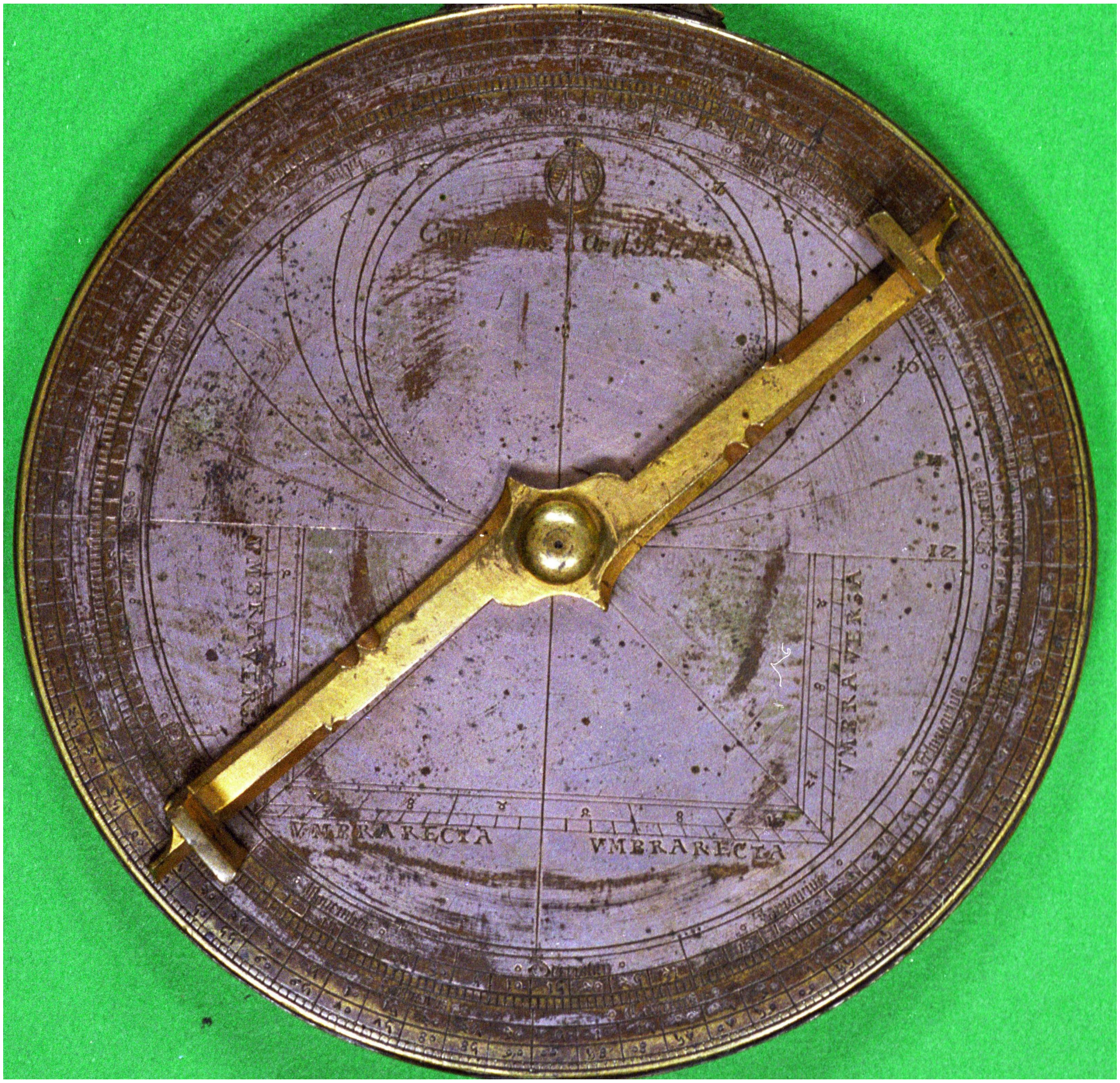

*Figure 7: The alidade mounted on the back of the mater*